\documentclass[5p,authoryear,preprint]{elsarticle}

\usepackage{pdflscape}

\usepackage{graphicx}
\usepackage{url}
\usepackage{amssymb}
\journal{Journal of High Energy Astrophysics}

\begin{document}

\begin{frontmatter}

%% Title, authors and addresses

%% use the tnoteref command within \title for footnotes;
%% use the tnotetext command for the associated footnote;
%% use the fnref command within \author or \address for footnotes;
%% use the fntext command for the associated footnote;
%% use the corref command within \author for corresponding author footnotes;
%% use the cortext command for the associated footnote;
%% use the ead command for the email address,
%% and the form \ead[url] for the home page:
%%
%% \title{Title\tnoteref{label1}}
%% \tnotetext[label1]{}
%% \author{Name\corref{cor1}\fnref{label2}}
%% \ead{email address}
%% \ead[url]{home page}
%% \fntext[label2]{}
%% \cortext[cor1]{}
%% \address{Address\fnref{label3}}
%% \fntext[label3]{}

\title{
Electron acceleration with improved Stochastic Differential Equation method:
cutoff shape of electron distribution in test-particle limit
}

%% use optional labels to link authors explicitly to addresses:
%% \author[label1,label2]{<author name>}
%% \address[label1]{<address>}
%% \address[label2]{<address>}

\author[aogaku]{Ryo~Yamazaki}
\author[ibaraki]{Tatsuo~Yoshida}
\author[aogaku]{Yuka~Tsuchihashi}
\author[aogaku]{Ryosuke~Nakajima}
\author[aogaku]{Yutaka~Ohira}
\author[ibaraki]{Shohei~Yanagita}

\address[aogaku]{
Department of Physics and Mathematics, Aoyama Gakuin University, 
5-10-1, Fuchinobe, Sagamihara 252-5258, Japan}
\address[ibaraki]
{College of Science, Ibaraki University, 2-1-1, Bunkyo, Mito 310-8512, Japan}

\begin{abstract}
We develop a method of stochastic differential equation
to simulate electron acceleration at astrophysical shocks.
Our method is based on It\^{o}'s stochastic differential equations coupled with a particle splitting,
employing a skew Brownian motion where an asymmetric shock crossing probability
is considered.
Using this code,
we perform simulations of electron acceleration at stationary plane
parallel shock with various parameter sets, and studied
how the cutoff shape, which is characterized by cutoff shape parameter $a$,
changes with the momentum dependence of the diffusion coefficient $\beta$.
In the age-limited cases, we reproduce previous results of other
authors, $a\approx2\beta$. 
In the cooling-limited cases,
the analytical expectation $a\approx\beta+1$ is roughly reproduced
although we recognize deviations to some extent.
In the case of escape-limited acceleration,
numerical result fits analytical stationary solution well, but
deviates from the previous asymptotic analytical formula 
$a\approx\beta$.

\end{abstract}


\end{frontmatter}

%%
%% Start line numbering here if you want
%%
% \linenumbers

%% main text

%%%%%%%%%%%%%%%%%%%%%%%%%%%%%%%%%%%%%%%%%%%%%%%%%%%%%%%%%%%%%%%%%%%%%%%%%%%

\section{Introduction}

Mechanism of particle acceleration is still unknown.
Diffusive shock acceleration \citep{krymskii77,bell1978,blandford1978}
is the most plausible
if strong shock waves exist  as in young supernova remnants (SNRs).
We have not yet well constrained model parameters, 
namely magnetic field strength and degree of magnetohydrodynamic 
turbulence, although there are observational claims of turbulent,
amplified field in young SNRs
\citep{vink03,bamba03,bamba05a,bamba05b,yamazaki04,uchiyama07}.
These are important to estimate maximum attainable energy of
both electrons and nuclei \citep[e.g.,][]{yoshida97}.
\citet{yamazaki13} proposed that cutoff shape of electron spectrum
around the maximum energy $E_{\rm max}$
may provide us important information on the cosmic-ray acceleration at
young SNRs.
They related
the cutoff shape parameter $a$, which is defined by 
$N(E)\propto\exp[-(E/E_{\rm max})^a]$,
to the energy dependence of the electron diffusion coefficient $\beta$
(that is, $K\propto E^\beta$)
in each case where the 
maximum electron energy
is determined by SNR age, synchrotron cooling and escape from the shock.
They found that if the power-law index of the electron spectrum
is independently determined by other observations, then 
the cutoff shape parameter can be constrained
by near future hard X-ray observations such as 
Nuclear Spectroscopic Telescope Array (NuSTAR) \citep{hailey10,harrison13}
and ASTRO-H \citep{takahashi10} and/or CTA \citep{actis11}.
These X-ray and gamma-ray observations will be important for the estimate 
of $\beta$ as well as $E_{\rm max}$ and the magnetic field strength.

In analysis of \citet{yamazaki13}, they assumed relations between
$a$ and $\beta$ as
%\begin{equation}
%a = \left\{
%\begin{array}{ll}
%2\beta    & ({\rm age-limited}) \\
%\beta+1   & ({\rm cooling-limited}) \\
%\beta     & ({\rm escape-limited}) \\
%\end{array}
%\right. ~~,
%\label{eq:relation_yama13}
%\end{equation}
$a=2\beta$, $\beta+1$ and $\beta$ in the case of
age-limited, cooling-limited and escape-limited acceleration, respectively.
The formula $a=2\beta$ in the age-limited case
has been based on numerical simulation
\citep{kato03,kang09}, while
the others are obtained analytically on the assumption of stationary state, 
and they are not yet confirmed numerically.
In this paper, we study the cutoff shape of the electron spectrum
by numerically solving the transport equation describing 
diffusive shock acceleration, and study whether the above
relations are right or not.

We use a numerical method for solving cosmic-ray transport equation
(so-called, diffusion-convection equation),
which was proposed by \citet{achterberg92}.
This method is
based on the equivalence between the Fokker-Planck equation and the
It\^{o} stochastic differential equation (SDE) \citep{gardiner83}.
Subsequent studies have followed for various situations
\citep{krulls94,yoshida94,marcowith99,marcowith10,schure10}.
It should be noted that
the SDE method has an advantage if the transport equation has to be
solved in multi-dimensions. In practice, the importance of upstream
inhomogeneity for understanding of cosmic-ray acceleration at supernova
remnants has been pointed out by various authors \citep[e.g.,][]{inoue12}. 
In this case, it is clear to consider the particle acceleration in
three dimensions.

The simple-minded application of the SDE method has problems 
in actual numerical integration.
First, $\delta$-functions appear in SDE if we apply it to the shock front,
where the background fluid velocity as well as the diffusion coefficient
have a sudden jump. In order to avoid this, the shock structure is
artificially smoothed \citep{achterberg92}. However, even in this case,
the time step has to be small enough for the simulated particles not to
miss the sharp gradient at the shock front, which significantly slows down the
simulation. Furthermore, in actual simulation time, approximation of the
smooth shock transition causes incorrect particle spectrum.
This difficulty was solved by \citet{zhang00} 
who used the skew Brownian motion \citep{harrison81}
which can be solved by a scaling method that eliminated the
$\delta$-functions in the SDE.
Other numerical schemes to resolve this problem have also been proposed
\citep{marcowith99,achterberg11}.
Second problem is that a large dynamic range in particle momentum
causes low statistical accuracy at large momenta.
This difficulty was also resolved by employing 
a particle splitting technique \citep{yoshida94}.

In this paper, we first attempt to perform simulations of electron
acceleration incorporating {\it both} methods of \citet{zhang00}  and
particle splitting.
Owing to newly developed code,
simulated spectra have cutoff shape accurate enough to be compared with
analytical formulae.
As a first step, we focus on the cases of one-dimensional plane shock.
Extended studies for more complicated cases such as time dependent free escape boundary,
nonuniform magnetic fields, and/or multi-dimensional systems
(including spherical shock geometry) are simple but
remained as future works.

%%%%%%%%%%%%%%%%%%%%%%%%%%%%%%%%%%%%%%%%%%%%%%%%%%%%%%%%%%%%%%%%%%%%%%%%%%%

\section{Basic Equations and Numerical Method}
\label{sec:method}

%%% BASIC EQUATIONS %%%

\subsection{Basic equations}

In this paper, we consider one-dimensional system, that is, all quantities
depend on the spatial coordinate $x$. 
The diffusion-convection equation with energy-loss process is given by
\begin{eqnarray}
&&
\frac{\partial f}{\partial t} + \frac{\partial }{\partial x}
\left(vf-K\frac{\partial f}{\partial x}\right) \nonumber \\
&& \qquad
+\frac{1}{p^2} \frac{\partial }{\partial p}
\left[
\left(-\frac{p}{3}\frac{d v}{d x}+\frac{dp}{dt}\right)
p^2f\right]=0~~,
\label{eq:dc}
\end{eqnarray}
where $f(x,p,t)$ is the distribution function for electrons, and
$p$ is the electron momentum.
Functions $v(x)$ and $K(x,p)$ are background velocity field and
the spatial diffusion coefficient of the electrons, respectively. 
In this paper, we consider the synchrotron cooling. 
Then, the loss term becomes
\begin{equation}
\frac{dp}{dt}=-\beta_{syn}\gamma p~~,
\label{eq:loss}
\end{equation}
where
\begin{equation}
\beta_{syn} = \frac{\sigma_{\rm T} B^2}{6\pi m_e c}
\label{eq:beta}~~,
\end{equation}
and $\gamma=\sqrt{(p/m_ec)^2+1}$ is the electrons' Lorentz factor,
and $B$ is the magnetic field.
Physical constants, $\sigma_{\rm T}$, $m_e$ and $c$ are
Thomson cross section, mass of electron and velocity of light,
respectively.

Introducing new quantities,
\begin{equation}
u = \ln\left(\frac{p}{m_ec}\right)~~,
\end{equation}
and
\begin{equation}
F(x,u,t)=p^3f(x,p,t)~~, 
\end{equation}
equation~(\ref{eq:dc}) becomes the Fokker-Planck form,
\begin{eqnarray}
&&
\frac{\partial F}{\partial t} 
+ \frac{\partial}{\partial x}\left[
\left(v+\frac{\partial K}{\partial x}\right)F\right]
- \frac{\partial^2}{\partial x^2}(KF) \nonumber\\
&& \qquad\qquad
-\frac{\partial}{\partial u}\left[
\left(\frac{1}{3}\frac{d v}{d x}+\beta_{syn}\gamma\right)F\right]=0
\label{eq:fp}~~.
\end{eqnarray}
This equation is equivalent to the following SDEs of the It\^{o} form:
\begin{equation}
dx = \left(v+\frac{\partial K}{\partial x}\right)dt+\sqrt{2K}dW~~,
\label{eq:sde1}
\end{equation}
\begin{equation}
du = -\left(\frac{1}{3}\frac{d v}{d x}+\beta_{syn}\gamma
       \right)dt~~,
\label{eq:sde2}
\end{equation}
where $dW$ is a Wiener process given by the Gaussian distribution:
\begin{equation}
P(dW)=\frac{1}{\sqrt{2\pi dt}}\exp(-dW^2/2dt)~~.
\end{equation}
Numerical simulation by SDEs is much faster than that with the usual
Monte Carlo method and is much easier than solving the original Fokker-Planck
equation, because the SDEs are ordinary differential equations.

%%% ZHANG's METHOD %%%

\subsection{Method of Zhang (2000)}

The application of the SDEs, equations~(\ref{eq:sde1}) and (\ref{eq:sde2}),
for the study of electron acceleration at the shock is not simple,
because the velocity field $v(x)$ has a sudden jump at the shock front,
so that $dv/dx$ in  equation~(\ref{eq:sde2}) contains $\delta$-function.
Similarly, if the diffusion coefficient also behaves discontinuously at
the shock front, then $\partial K/\partial x$ in  equation~(\ref{eq:sde1}) 
also contains  the $\delta$-function.
We take the comoving frame with the shock which is located at $x=0$
and we define $x<0$ as upstream region.
Following \citet{zhang00}, we decompose the velocity field $v$
and the diffusion coefficient $K$ into two parts:
\begin{equation}
v(x) = v_c(x) + \frac{\Delta V}{2}{\rm sign}(x)~~,
\end{equation}
\begin{equation}
K(x) = K_c(x) + \frac{\Delta K}{2}{\rm sign}(x)~~,
\end{equation}
where $\Delta V=v(0^+)-v(0^-)$ and $\Delta K=K(0^+)-K(0^-)$,
and ${\rm sign}(x)$ is the sign of $x$.
Functions $v_c(x)$ and $K_c(x)$ are continuous for arbitrary $x$
(including $x=0$).
We scale the $x$ coordinate according to its sign in the following way
\citep{harrison81}:
\begin{equation}
y=x s(x) = x\times \left\{
\begin{array}{cc}
\alpha & (x<0) \\
\frac{1}{2} & (x=0) \\
1-\alpha & (x>0) \\
\end{array}
\right. ~~,
\label{eq:s}
\end{equation}
where
\begin{equation}
\alpha = \frac{K(0^+)}{K(0^+)+K(0^-)}~~.
\label{eq:alpha}
\end{equation}
Then, SDEs (\ref{eq:sde1}) and (\ref{eq:sde2}) can be rewritten as
\begin{equation}
dy = s(x) \left[
\left( v(x) + \frac{\partial K_c}{\partial x} \right)dt + \sqrt{2K}dW\right]~~,
\label{eq:sde3} 
\end{equation}
\begin{equation}
du = -\left(\frac{1}{3}\frac{dv_c}{dx}+\beta_{syn}\gamma \right) dt
-\frac{\Delta V}{3\Delta K}[dx-s^{-1}(y)dy] ~~.
\label{eq:sde4}
\end{equation}
Derivation of equations~(\ref{eq:sde3}) and (\ref{eq:sde4}) are the
same way as of \citet{zhang00}.
These equations do not contain $\delta$-functions and can be integrated directly. 
Once $y(t)$ is obtained, the position of electrons $x(t)$
can be obtained by
\begin{equation}
x = y s^{-1}(y) = y\times \left\{
\begin{array}{cc}
1/\alpha & (y<0) \\
2 & (y=0) \\
1/(1-\alpha) & (y>0) \\
\end{array}
\right. ~~.
\label{eq:sinv}
\end{equation}
In order to see the effect of diffusion, the spatial step size of
diffusion in one time step $\Delta t$ must be larger than that of
convection, that is $v\Delta t<\sqrt{2K\Delta t}$. Hence we derive
the requirement of the time step as
\begin{equation}
\Delta t < \frac{2K}{v^2}~~.
\label{eq:timestep}
\end{equation}

Functions $x(t)$ and $u(t)$ are numerically integrated as follows.
We define $X_i=X(t_i)$ and $X_{i+1}=X(t_i+\Delta t)$, where
$X=x$, $y$ and $u$.
First, we discretize equation~(\ref{eq:sde3}) as
\begin{eqnarray}
y_{i+1}&=&y_i(x_i) + s(x_i) \left[
\left( v(x_i) + \frac{\partial K_c}{\partial x}(x_i,u_i) \right)\Delta t
\right. 
\nonumber \\
&& \qquad\qquad
+ \sqrt{2K(x_i,u_i)}\Delta W\Bigr]~~,
\end{eqnarray}
where $\Delta W$ is independent and identically distributed normal random 
variable with expected value zero and variance $\Delta t$.
Then, $y_{i+1}$ is obtained for given $x_{i}$ and $u_{i}$.
Second, we get $x_{i+1}$ from equation~(\ref{eq:sinv}).
Finally, $u_{i+1}$ is calculated from equation~(\ref{eq:sde4}),
%% \begin{equation}
%% D=-\frac{\Delta V}{3\Delta K}[dx-s^{-1}(y)dy]~~,
%% \end{equation}
which is discretized as
\begin{equation}
u_{i+1} = u_i
-\left(\frac{1}{3}\frac{dv_c}{dx}(x_i)+\beta_{syn}\gamma(u_i) \right) \Delta t
+\Delta L~~,
\end{equation}
where
\begin{eqnarray}
\Delta L 
  &=& -\frac{\Delta V}{3\Delta K}
       [(x_{i+1}-x_i)-s^{-1}(y_i)(y_{i+1}-y_i)]\nonumber\\
  &=& -\frac{\Delta V}{3\Delta K}
       [x_{i+1}-s^{-1}(y_i)y_{i+1}] \nonumber\\
  &=& \frac{-\Delta V}{3[K(0^+)+K(0^-)]} \nonumber\\
  && 
   \times\left\{
   \begin{array}{cc}
      0 & ( x_i x_{i+1}>0 ) \\
%      \frac{x_{i+1}}{\alpha-1} ~~~ (x_i>0 ~~{\rm and}~~x_{i+1}<0 ) \\
%      \frac{x_{i+1}}{\alpha} ~~~ (x_i<0 ~~{\rm and}~~x_{i+1}>0) \\
      x_{i+1}/(\alpha-1) & (x_i>0, x_{i+1}<0 ) \\
      x_{i+1}/\alpha & (x_i<0, x_{i+1}>0) \\
      |x_{i+1}| & ( x_i=0, x_{i+1}\ne0 ) \\
      0 & (x_{i+1}=0 ) \\
      \end{array}
  \right. ~~.
\end{eqnarray}
Here we use the fact $0<\alpha<1$ and equations~(\ref{eq:s}) and (\ref{eq:sinv}).
Now the meaning of $\Delta L$ becomes clear.
Since $\Delta V<0$ in our case (see section~\ref{sec:setup}),
one can find $\Delta L>0$ when $x_ix_{i+1}<0$, and $\Delta L=0$ 
when $x_ix_{i+1}>0$.
Therefore, it is confirmed that particles gain energy when 
they pass through the shock front \citep{zhang00}.

%%% PARTICLE SPLITTING %%%

\subsection{Particle Splitting}

A particle splitting is necessary to achieve a wide momentum range of 
accelerated electrons.
Following \citet{yoshida94},
we set splitting surfaces $u_n$ in momentum space 
($u_{\rm s0}<u<u_{\rm s1}$)
with an equal spacing in logarithmic scale:
\begin{equation}
u_n=u_{\rm s0} + n\Delta u~~,
\end{equation}
where $n=0,1,2,\cdots,n_{\rm max}$ and 
$\Delta u = (u_{\rm s1}-u_{\rm s0})/n_{\rm max}$.
Each time an accelerated particle hits the surface $u_n$, the particle is
split into $w$ particles with the same energy and spatial
position which particles have attained.
The statistical weight which is needed to calculate the final spectrum of the particles
is decreased by a factor of $w$ in each splitting.
%

%%%%%%%%%%%%%%%%%%%%%%%%%%%%%%%%%%%%%%%%%%%%%%%%%%%%%%%%%%%%%%%%%%%%%%%%%%%

\section{Simulations for stationary shock cases}

\subsection{Simulation setup}
\label{sec:setup}

In the following, we consider electron acceleration at 
stationary plane parallel shock, that is,
\begin{equation}
v_c(x) = \frac{v_1+v_2}{2}~~,
\end{equation}
and $\Delta V = v_2-v_1$, so that
\begin{equation}
v(x) =  \left\{
\begin{array}{cc}
v_1 & (x<0) \\
v_2 & (x>0) \\
\end{array}
\right. ~~,
\label{eq:vel3}
\end{equation}
where constants $v_1$ and $v_2$ are upstream and downstream velocities, respectively.
Compression ratio $r=v_1/v_2$ is fixed to be 4 throughout the paper.
We also assume that the diffusion coefficient is uniform both in upstream 
and downstream regions, that is,
\begin{equation}
K_c(x) = \frac{K_1+K_2}{2}~~,
\end{equation}
and $\Delta K = K_2-K_1$, so that
\begin{equation}
K(x) =  \left\{
\begin{array}{cc}
K_1 & (x<0) \\
K_2 & (x>0) \\
\end{array}
\right. ~~,
\label{eq:diff3}
\end{equation}
where upstream and downstream coefficients, $K_1$ and $K_2$, depend on
electron momentum. In this paper, we assume
\begin{eqnarray}
K_1(p) &=& rK_2(p) \nonumber\\
    &=&1.6\times10^{19} B_{\mu {\rm G}}^{-1}
           \left(\frac{p}{m_ec}\right)^\beta
           {\rm cm}^2{\rm s}^{-1} ~~,
\label{eq:diff4}
\end{eqnarray}
where $B_{\mu {\rm G}}$ is the magnetic field strength in units of $\mu$G.

We set a free escape boundary at $x=-x_{\rm feb}$ ($<0$) 
in the upstream region. Once a particle goes beyond the boundary, it
never comes back to the acceleration site.
This fact becomes significant when the particle's penetration depth 
in the upstream region,
$K_1(p)/v_1$, is comparable to $x_{\rm feb}$.
On the other hand, when $x_{\rm feb}$ is sufficiently large 
(i.e., $x_{\rm feb}\rightarrow\infty$), the particle escape does not occur.

The start and end times of electron acceleration are $t=0$ and
$t_{\rm age}$, respectively. During this period, electrons are
injected with a momentum $p_{\rm inj}$ at the shock front $x=0$
continuously at a constant rate.
Taking an ensemble average over a number of realizations of SDEs,
(\ref{eq:sde3}), (\ref{eq:sde4}) and (\ref{eq:sinv}),
we obtain the momentum spectrum of the whole region
(including both upstream and downstream regions as well as the shock front) 
at $t=t_{\rm age}$.
In the following, we consider the case $\beta>0$.
Then, $K(p)$ increases with $p$.
Hence, if the injection momentum $p_{\rm inj}$ is taken so as to
$\Delta t < 2K(p_{\rm inj})/v_1^2$,
the condition for time step, equation~(\ref{eq:timestep}),
is always satisfied. 
It can be seen from Table~2 that
$p_{\rm inj}$ satisfies this requirement.

We perform simulations for various parameter sets. 
Adopted parameters are summarized in Tables~1 and 2.
The cutoff shape of electron spectrum
depends on how the maximum momentum of electrons
is determined. In the next section, we consider three cases,
age-limited, cooling-limited, and escape-limited cases, in order to
decide the maximum attainable electron momentum due to the 
diffusive shock acceleration.

\begin{table*}
\label{table1}
\centering
\begin{minipage}{140mm}
\caption{Adopted parameters in the present study.}
\begin{tabular}{@{}ccrcrrcccc@{}}
\hline
Run\footnote{A, C and E stand for age, cooling and escape, respectively.} & 
~~~~$\beta$~~~~ &    $B$        &    $v_1$      & 
$t_{\rm age}$   & $x_{\rm feb}$ & $p_{\rm inj}$ & 
$p_{\rm m,age}$\footnote{calculated according to equation~(\ref{eq:Emaxage}).}  & 
$p_{\rm m,cool}$\footnote{calculated according to equation~(\ref{eq:Emaxcool}).} & 
$p_{\rm m,esc}$\footnote{calculated according to equation~(\ref{eq:Emaxesc}).} \\
& & [$\mu$G] & [$10^8$cm~s$^{-1}$] & [yr] & [$10^{15}$cm] & 
[$m_ec$] & [$m_ec$] & [$m_ec$] & [$m_ec$] \\
\hline
\hline
A07-1  & 0.7 &    1 & 1 &  3  &   $\infty$ & $10^3$  
                 &  $3.4\times10^5$  &  $7.4\times10^9$  &     $\infty$     \\
A07-2  & 0.7 &    1 & 1 & 10  &   $\infty$ & $10^3$  
                 &  $1.9\times10^6$  &  $7.4\times10^9$  &     $\infty$     \\ 
A07-3  & 0.7 &    1 & 1 & 30  &   $\infty$ & $10^3$  
                 &  $9.0\times10^6$  &  $7.4\times10^9$  &     $\infty$     \\ 
A07-4  & 0.7 &    1 & 1 &100  &   $\infty$ & $10^3$  
                 &  $5.0\times10^7$  &  $7.4\times10^9$  &     $\infty$     \\ 
A07-5  & 0.7 &    1 & 1 &300  &   $\infty$ & $10^3$  
                 &  $2.4\times10^8$  &  $7.4\times10^9$  &     $\infty$     \\  
\hline
A10-1  & 1.0 &    1 & 6 &  3  &   $\infty$ & $10^3$  
                 &  $2.7\times10^5$  &  $1.5\times10^9$  &     $\infty$     \\
A10-2  & 1.0 &    1 & 6 & 10  &   $\infty$ & $10^3$  
                 &  $8.9\times10^5$  &  $1.5\times10^9$  &     $\infty$     \\ 
A10-3  & 1.0 &    1 & 6 & 30  &   $\infty$ & $10^3$  
                 &  $2.7\times10^6$  &  $1.5\times10^9$  &     $\infty$     \\ 
A10-4  & 1.0 &    1 & 6 &100  &   $\infty$ & $10^3$  
                 &  $8.9\times10^6$  &  $1.5\times10^9$  &     $\infty$     \\ 
A10-5  & 1.0 &    1 & 6 &300  &   $\infty$ & $10^3$  
                 &  $2.7\times10^7$  &  $1.5\times10^9$  &     $\infty$     \\ 
\hline
A15-1  & 1.5 &    5 & 8 &  3  &   $\infty$ & $10^3$  
                 &  $1.8\times10^4$  &  $1.4\times10^7$  &     $\infty$     \\
A15-2  & 1.5 &    5 & 8 & 10  &   $\infty$ & $10^3$  
                 &  $4.0\times10^4$  &  $1.4\times10^7$  &     $\infty$     \\ 
A15-3  & 1.5 &    5 & 8 & 30  &   $\infty$ & $10^3$  
                 &  $8.2\times10^4$  &  $1.4\times10^7$  &     $\infty$     \\ 
A15-4  & 1.5 &    5 & 8 &100  &   $\infty$ & $10^3$  
                 &  $1.8\times10^5$  &  $1.4\times10^7$  &     $\infty$     \\ 
A15-5  & 1.5 &    5 & 8 &300  &   $\infty$ & $10^3$  
                 &  $3.8\times10^5$  &  $1.4\times10^7$  &     $\infty$     \\ 
\hline
\hline
C07-1  & 0.7 & 2000 & 0.1 & 10  &   $\infty$ & $10^4$  
                 &  $1.4\times10^8$  &  $5.7\times10^6$  &     $\infty$     \\ 
C07-2  & 0.7 &  500 & 0.1 & 100 &   $\infty$ & $10^4$  
                 &  $5.0\times10^8$  &  $1.3\times10^7$  &     $\infty$     \\ 
\hline
C10-1  & 1.0 & 2000 & 1   & 10  &   $\infty$ & $10^4$  
                 &  $4.9\times10^7$  &  $5.5\times10^6$  &     $\infty$     \\ 
C10-2  & 1.0 &  500 & 1   & 100 &   $\infty$ & $10^4$  
                 &  $1.2\times10^8$  &  $1.1\times10^7$  &     $\infty$     \\ 
\hline
C15-1  & 1.5 & 2000 & 10  & 100 &   $\infty$ & $10^4$  
                 &  $1.3\times10^7$  &  $1.6\times10^6$  &     $\infty$     \\ 
C15-2  & 1.5 &  500 & 10  & 800 &   $\infty$ & $10^4$  
                 &  $2.1\times10^7$  &  $2.7\times10^6$  &     $\infty$     \\ 
\hline
\hline
E07-1  & 0.7 &    1 & 6 &  95  &  0.1 & $10^2$  
                 &  $7.8\times10^9$  & $6.1\times10^{10}$ &  $1.3\times10^5$ \\
E07-2  & 0.7 &    1 & 6 &  95  &   1  & $10^2$  
                 &  $7.8\times10^9$  & $6.1\times10^{10}$ &  $3.4\times10^6$ \\
E07-3  & 0.7 &    1 & 6 &  95  &   10 & $10^2$  
                 &  $7.8\times10^9$  & $6.1\times10^{10}$ &  $9.2\times10^7$ \\
E07-4  & 0.7 &    1 & 6 &  95  &  100 & $10^2$  
                 &  $7.8\times10^9$  & $6.1\times10^{10}$ &  $2.5\times10^9$ \\
\hline
E10-1  & 1.0 &    1 & 6 &  95  &  0.1 & $10^2$  
                 &  $8.4\times10^6$  &  $1.5\times10^9$  &  $3.8\times10^3$ \\
E10-2  & 1.0 &    1 & 6 &  95  &   1  & $10^2$  
                 &  $8.4\times10^6$  &  $1.5\times10^9$  &  $3.8\times10^4$ \\
E10-3  & 1.0 &    1 & 6 &  95  &   10 & $10^2$  
                 &  $8.4\times10^6$  &  $1.5\times10^9$  &  $3.8\times10^5$ \\
E10-4  & 1.0 &    1 & 6 &  95  &  100 & $10^2$  
                 &  $8.4\times10^6$  &  $1.5\times10^9$  &  $3.8\times10^6$ \\
\hline
E15-1  & 1.5 &    1 & 6 &  95  &  0.1 & $10^2$  
                 &  $4.1\times10^4$  &  $2.2\times10^7$  &  $2.4\times10^2$ \\
E15-2  & 1.5 &    1 & 6 &  95  &   1  & $10^2$  
                 &  $4.1\times10^4$  &  $2.2\times10^7$  &  $1.1\times10^3$ \\
E15-3  & 1.5 &    1 & 6 &  95  &   10 & $10^2$  
                 &  $4.1\times10^4$  &  $2.2\times10^7$  &  $5.2\times10^3$ \\
E15-4  & 1.5 &    1 & 6 &  95  &  100 & $10^2$  
                 &  $4.1\times10^4$  &  $2.2\times10^7$  &  $2.4\times10^4$ \\
\hline
\end{tabular}
\end{minipage}
\end{table*}

%%%%%%%%%%%%%%%%%%%%%%%%%%%%%%%%%%%%%%%%%%%%%%%%%%%%%%%%%%%%%%%%%%%%%%%%%%%

\subsection{Estimate of Maximum electron momentum}
\label{sec:Emax}
 
The maximum momentum of accelerated electrons is limited by 
a finite shock age, their cooling or escape \citep[e.g.,][]{yamazaki06,ohira12}.
It is obtained by comparisons of timescales, 
which are given as functions of electron momentum 
and the shock age, $t_{\rm age}$ in the age-limited and cooling-limited cases. 
The acceleration time of the
diffusive shock acceleration is represented by \citep{drury83}
\begin{equation}
t_{\rm acc} = \frac{3}{v_1-v_2}\left(
\frac{K_1}{v_1}+\frac{K_2}{v_2}\right)~~.
\label{eq:acc1}
\end{equation}
Using $v_2=v_1/r$ and equation~(\ref{eq:diff4}) with $r=4$, we obtain
\begin{equation}
t_{\rm acc}=1.28\times10^4 B_{\mu {\rm G}}^{-1} v_8^{-2}
\left(\frac{p}{m_ec}\right)^\beta {\rm s}^{-1} ~~,
\label{eq:acc2}
\end{equation}
where $v_8$ is the shock velocity, $v_1$, in units of $10^8$cm~s$^{-1}$.
From the condition $t_{\rm acc}=t_{\rm age}$,
the age-limited maximum momentum, $p_{\rm m,age}$, is derived as
\begin{equation}
p_{\rm m,age}=
(2.46\times10^5B_{\mu {\rm G}}v_8^2t_{100})^{\frac{1}{\beta}}m_ec ~~,
\label{eq:Emaxage}
\end{equation}
where $t_{100}=t_{\rm age}/100~{\rm yr}$.

When the magnetic field is strong, 
the electron acceleration is limited by synchrotron cooling. 
We obtain the cooling-limited maximum momentum from
the condition $t_{\rm acc}=t_{\rm cool}$, where 
$t_{\rm cool}$ is the synchrotron cooling time given by
\begin{equation}
t_{\rm cool} = 
%%\frac{6\pi m_ec}{\sigma_{\rm T}B^2}
\beta_{\rm syn}^{-1}
\left( \frac{p}{m_ec}\right)^{-1}~~.
\label{eq:cool}
\end{equation}
Using equations~(\ref{eq:acc2}) and (\ref{eq:cool}), we derive
\begin{equation}
p_{\rm m,cool} =
(6.05\times10^{16}B_{\mu {\rm G}}^{-1}v_8^2)^{\frac{1}{\beta+1}}m_ec~~.
\label{eq:Emaxcool}
\end{equation}

It may happen that the maximum energy is limited by the escape process
\citep{ohira10}.
Characteristic spatial length of particles
penetrating into the upstream region is given by $K_1(p)/v_1$.
As long as $K_1(p)/v_1\ll x_{\rm feb}$, the particles are confined without the
significant escape loss, and they are accelerated to higher energies.
On the other hand, when their momentum increases up to
sufficiently high energies satisfying $K_1(p)/v_1> x_{\rm feb}$,
their acceleration ceases and they escape into the far upstream. 
Therefore, the
maximum momentum of accelerated particles in this scenario is
given by the condition 
\begin{equation}
\frac{K_1(p)}{v_1}=x_{\rm feb}~~.
\label{eq:esc}
\end{equation}
This reads
\begin{equation}
p_{\rm m,esc} =
(6.25\times10^3 B_{\mu {\rm G}}v_8x_{15})^{\frac{1}{\beta}}m_ec~~,
\label{eq:Emaxesc}
\end{equation}
where $x_{15}=x_{\rm feb}/10^{15}{\rm cm}$.

When $p_{\rm m,age}$ is smallest among $p_{\rm m,age}$,
$p_{\rm m,cool}$ and $p_{\rm m,esc}$, the acceleration is limited by 
the age of the shock. In the cooling-limited and escape-limited cases,
$p_{\rm m,cool}$ and $p_{\rm m,esc}$ are smallest, respectively.

%%%%%%%%%%%%%%%%%%%%%%%%%%%%%%%%%%%%%%%%%%%%%%%%%%%%%%%%%%%%%%%%%%%%%%%%%%%

\subsection{Results}
\label{sec:results}

\begin{figure*}
\begin{center}
\includegraphics[width=55mm]{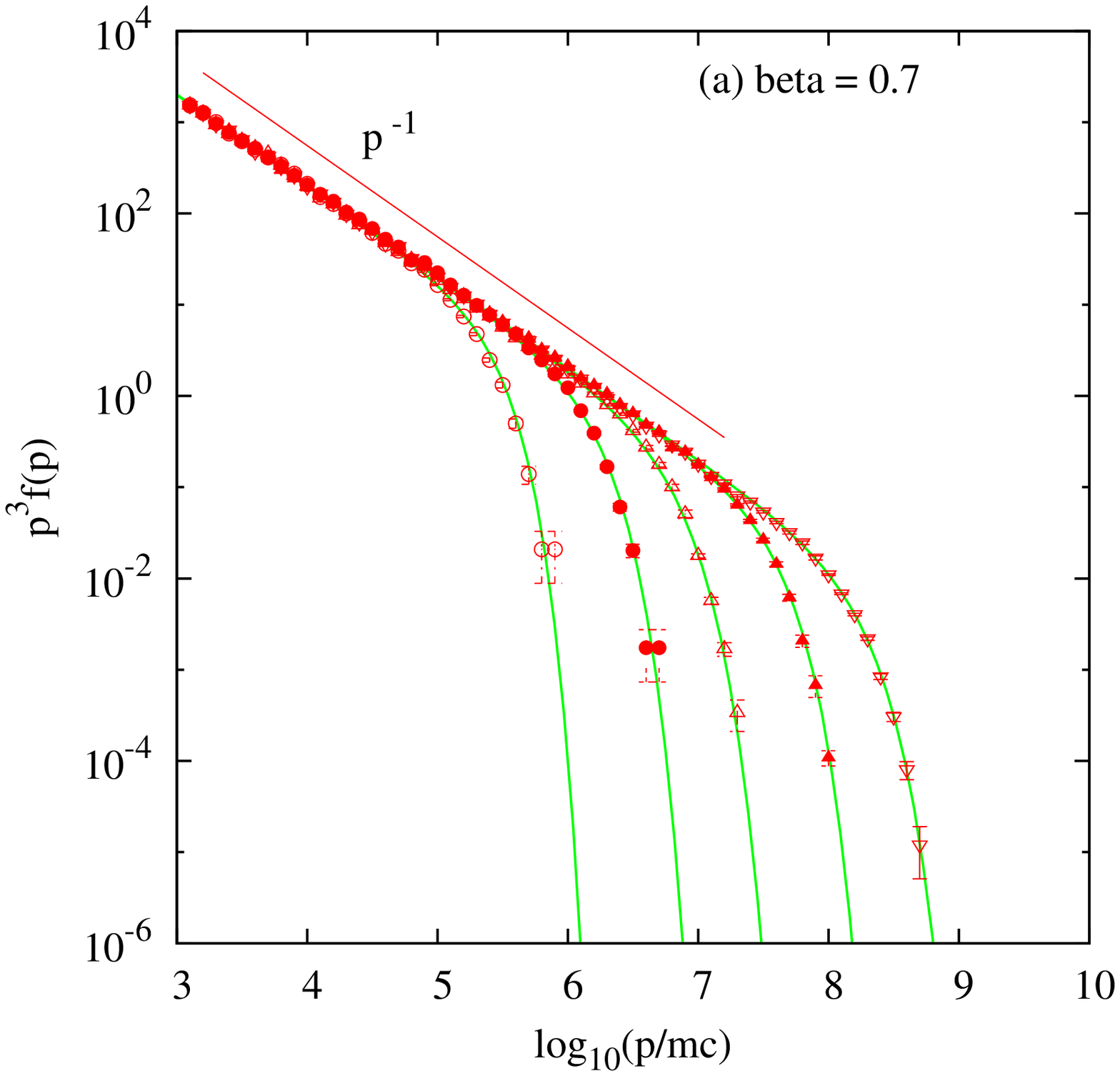}
\includegraphics[width=55mm]{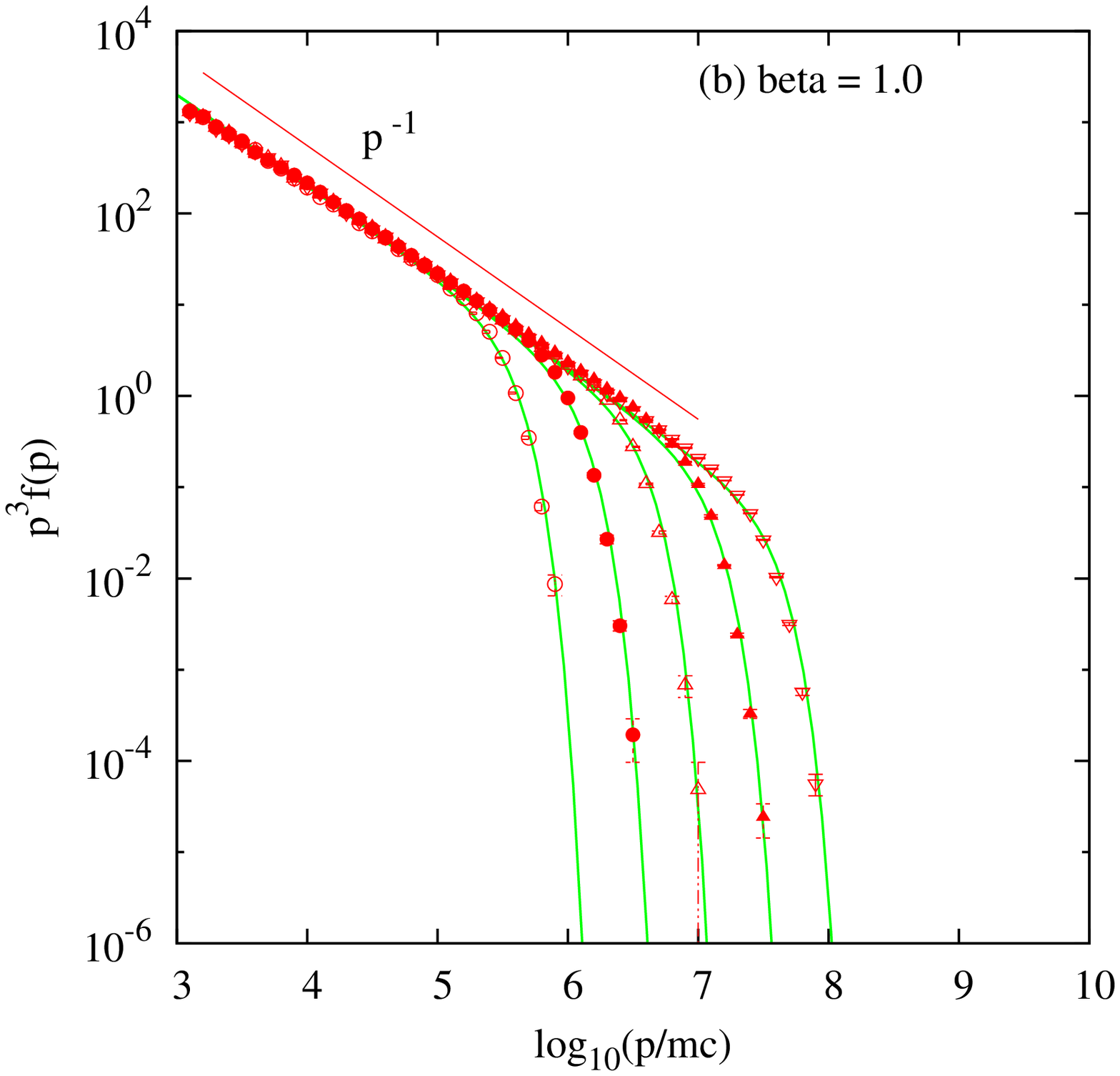}
\includegraphics[width=55mm]{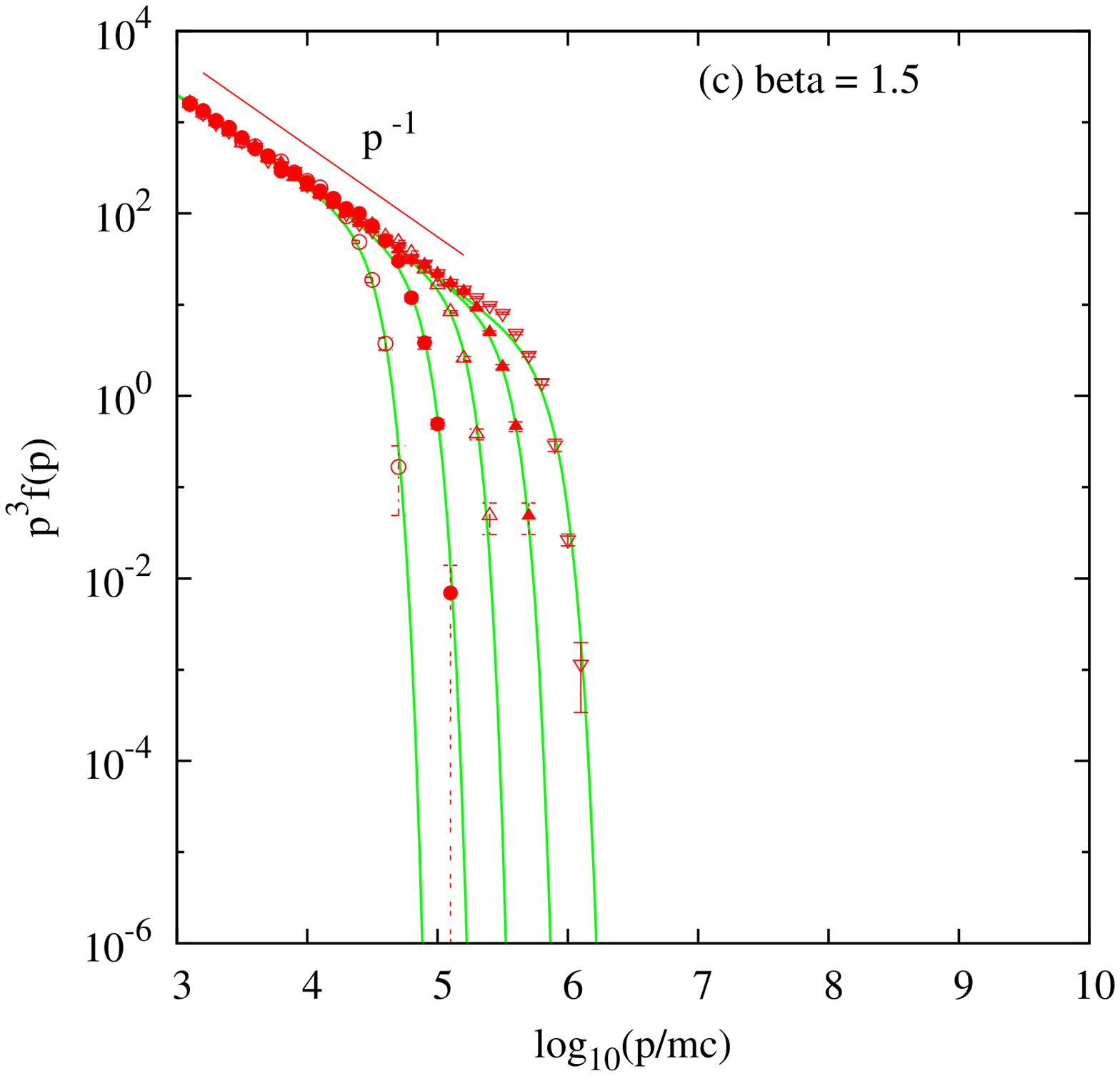}
\end{center}
\caption{
Electron spectra in the age-limited cases with
(a) $\beta=0.7$ (Runs A07-1, A07-2, A07-3, A07-4, A07-5 from left to right),
(b) $\beta=1.0$ (Runs A10-1, A10-2, A10-3, A10-4, A10-5 from left to right)
and
(c) $\beta=1.5$ (Runs A15-1, A15-2, A15-3, A15-4, A15-5 from left to right).
Lines indicate the best fitted models described by equation~(\ref{eq:shape_age}).
}
\label{fig:spec_age}
\end{figure*}

\begin{figure*}
\begin{center}
\includegraphics[width=55mm]{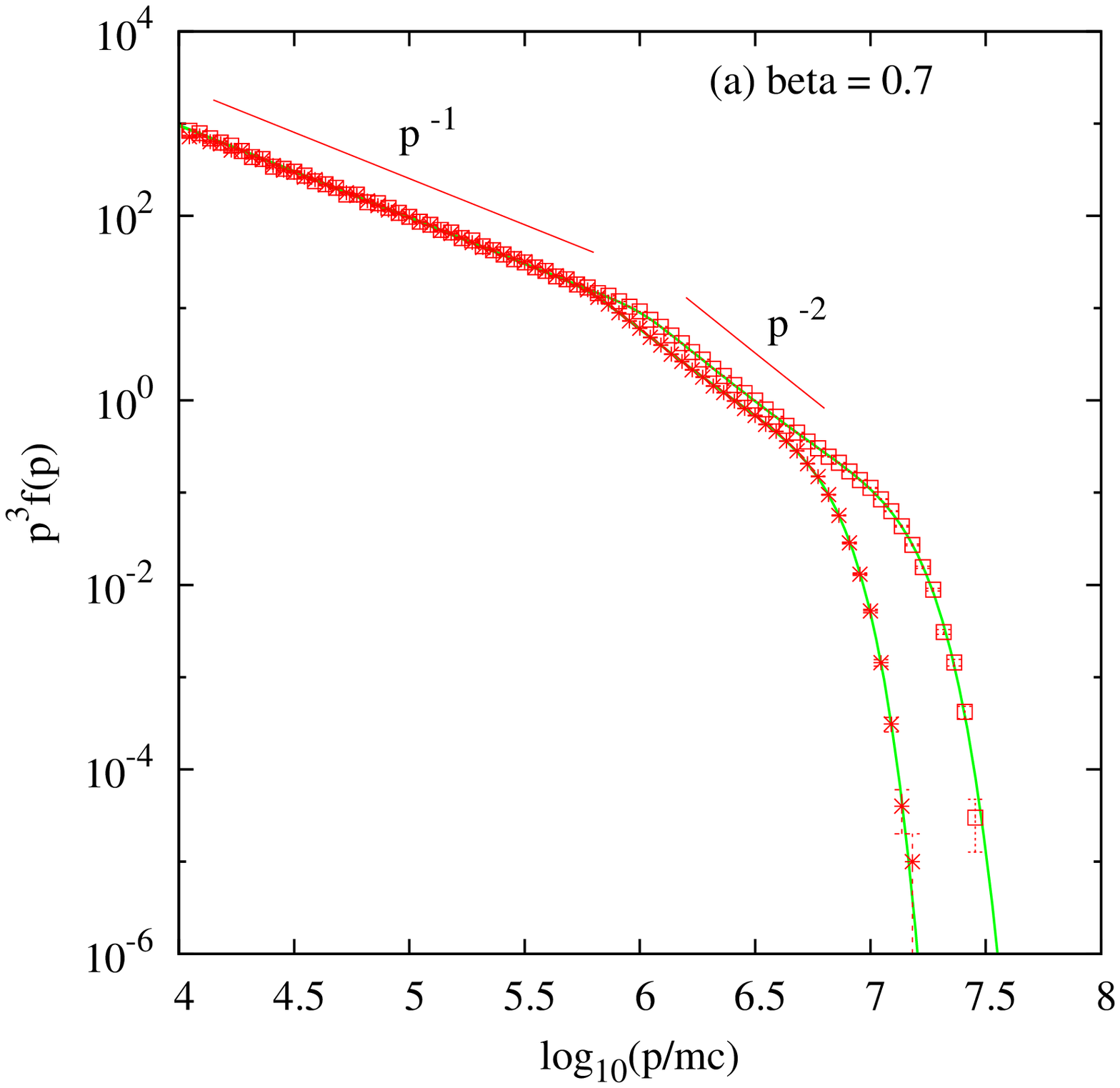}
\includegraphics[width=55mm]{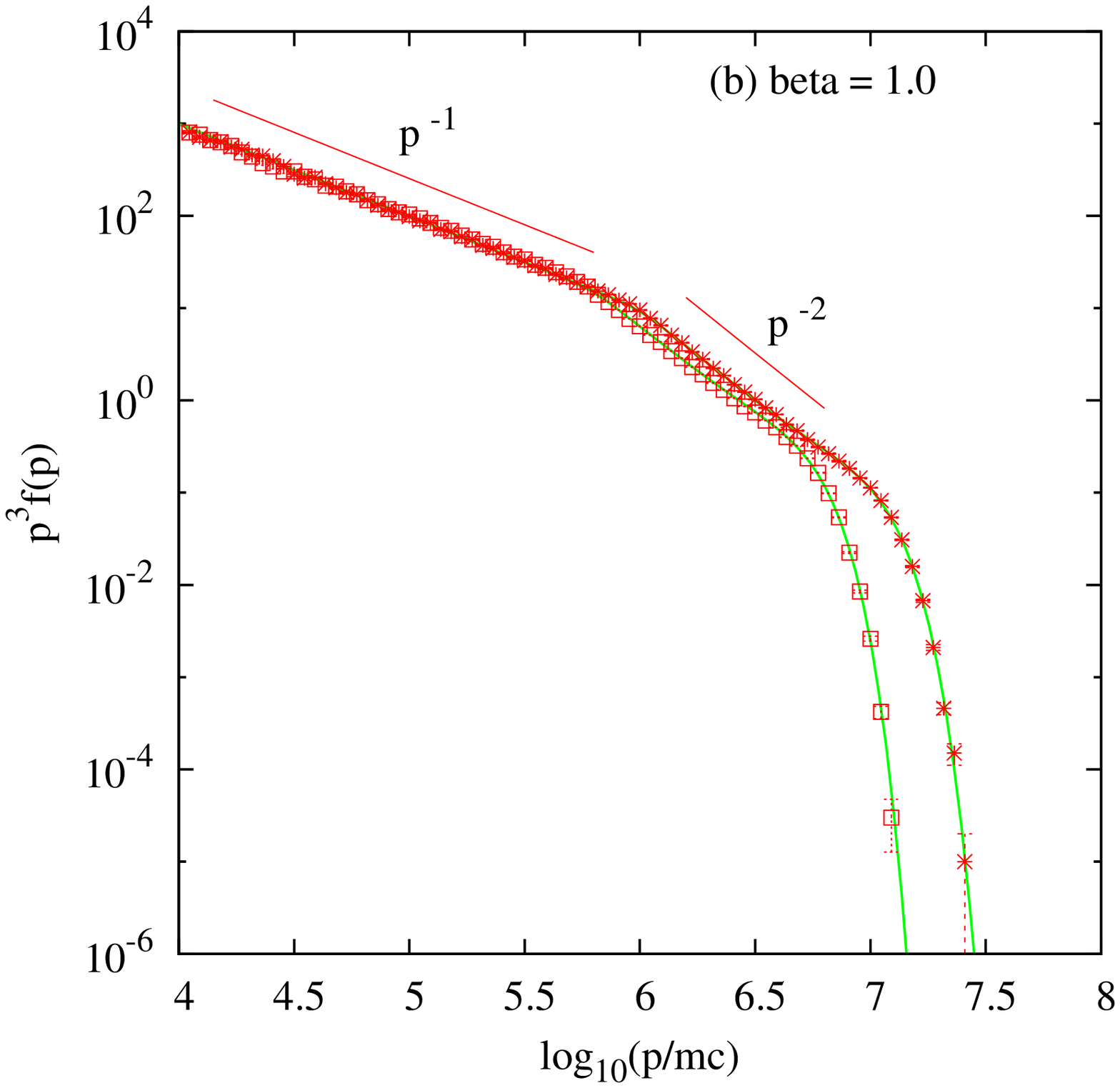}
\includegraphics[width=55mm]{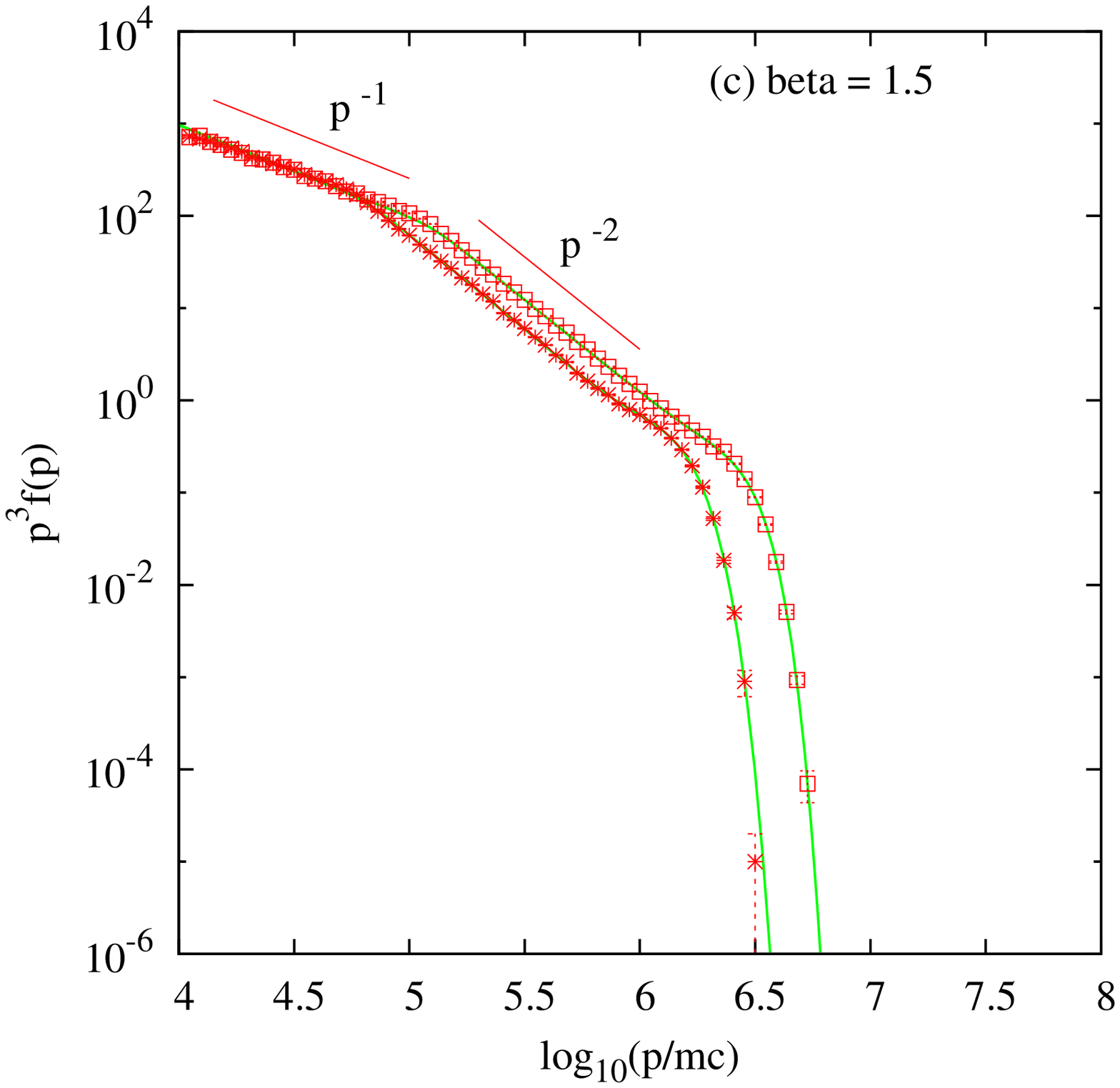}
\end{center}
\caption{
Electron spectra in the cooling-limited cases with
(a) $\beta=0.7$ (Runs C07-1, C07-2 from left to right),
(b) $\beta=1.0$ (Runs C10-1, C10-2 from left to right)
and
(c) $\beta=1.5$ (Runs C15-1, C15-2 from left to right).
Lines indicate the best fitted models described by equations~(\ref{eq:fit_cool}),
(\ref{eq:fit_fb}) and (\ref{eq:fit_fp}).
}
\label{fig:spec_cool}
\end{figure*}

\begin{figure*}
\begin{center}
\includegraphics[width=55mm]{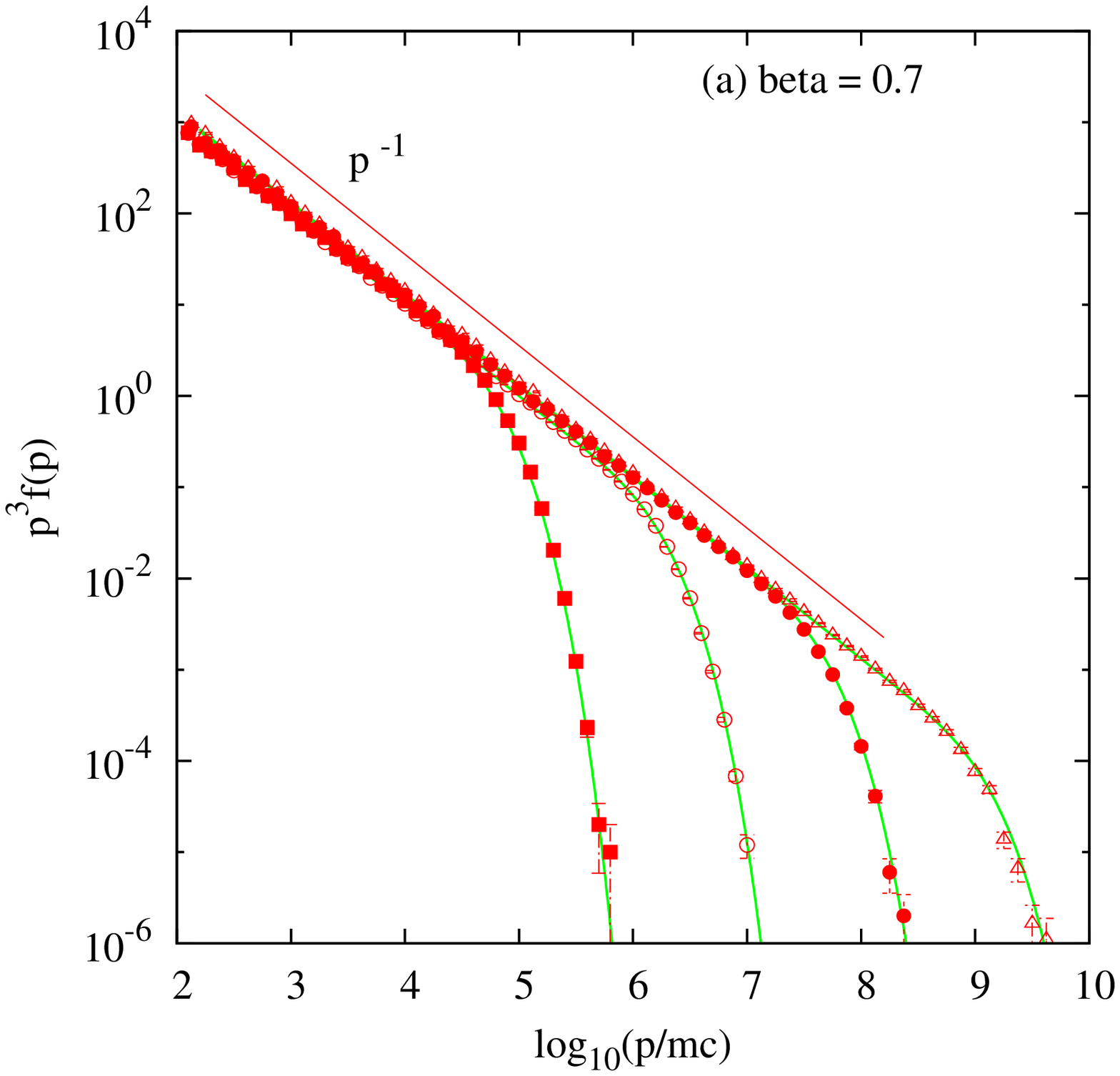}
\includegraphics[width=55mm]{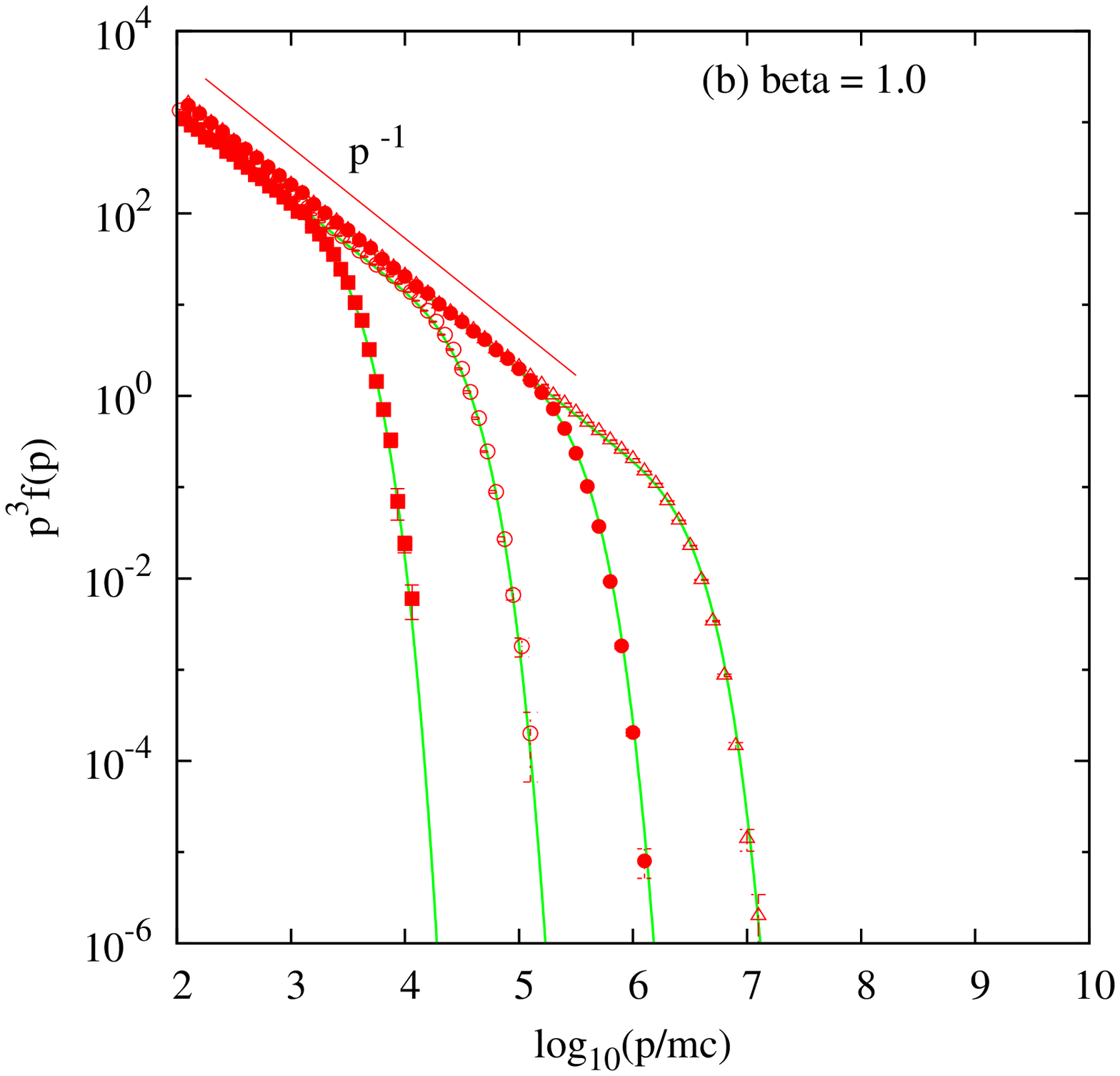}
\includegraphics[width=55mm]{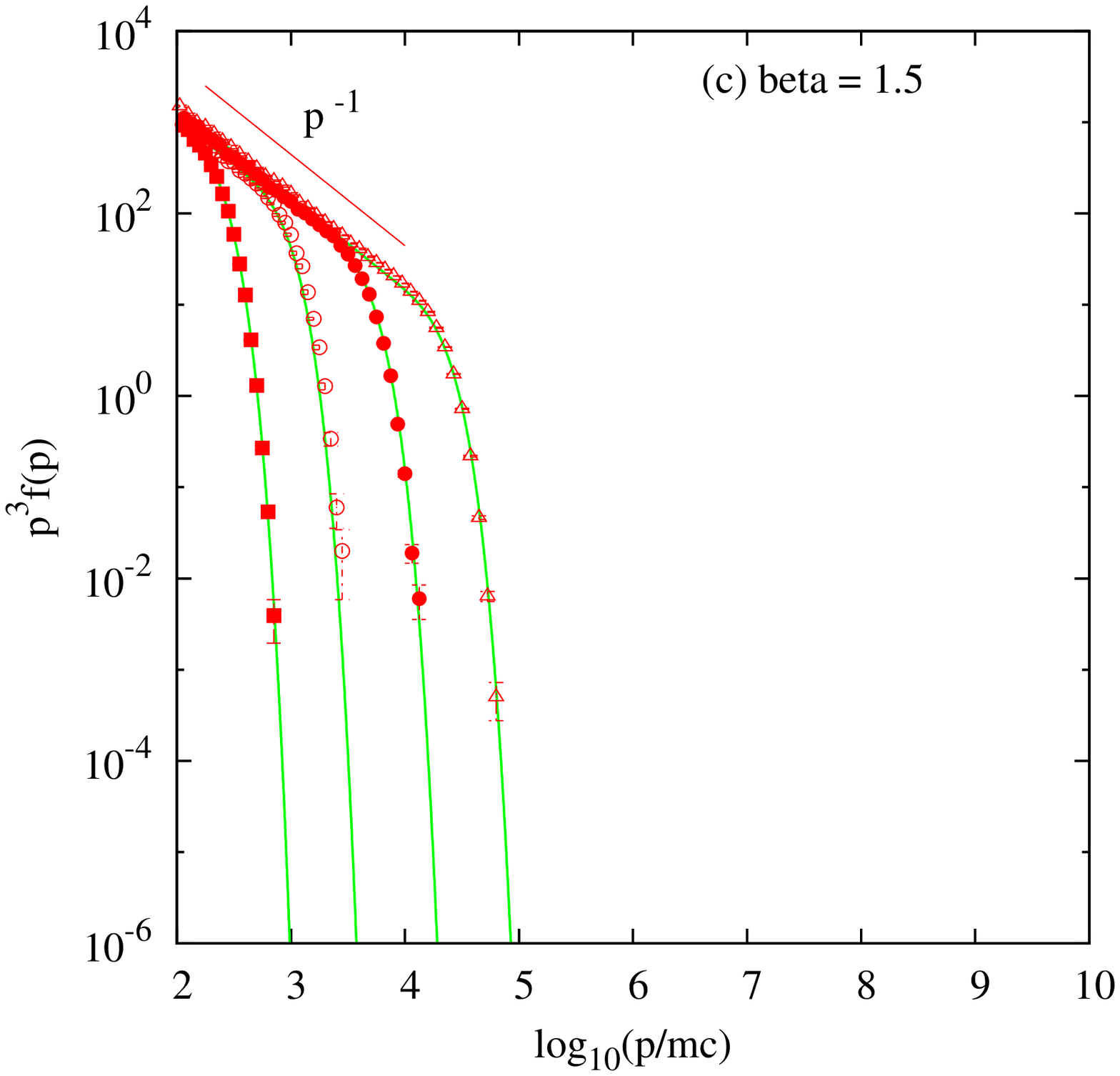}
\end{center}
\caption{
Electron spectra in the escape-limited cases with
(a) $\beta=0.7$ (Runs E07-1, E07-2, E07-3, E07-4 from left to right),
(b) $\beta=1.0$ (Runs E10-1, E10-2, E10-3, E10-4 from left to right)
and
(c) $\beta=1.5$ (Runs E15-1, E15-2, E15-3, E15-4 from left to right).
Lines indicate the best fitted models described by equation~(\ref{eq:anal_esc}).
}
\label{fig:spec_esc}
\end{figure*}

Figures~\ref{fig:spec_age}, \ref{fig:spec_cool} and \ref{fig:spec_esc} 
show results of numerical simulation for the age-limited,
cooling-limited and escape-limited cases, respectively.
The spectra in these figures are for all particles which
are still in the system at $t_{\rm age}$, that is,
\begin{equation}
F(p) \propto
p^3 f(p) \propto
\int_{-x_{\rm feb}}^{\infty}F(x,u,t_{\rm age})\,dx ~~.
\end{equation} 
The value $F$ of the distribution function for given momentum range $[u,u+\Delta u]$ 
is derived by an ensemble average.
If injected particles with momentum $p_{\rm inj}$
have a statistical weight of unity, 
then, particles which have experienced splitting $n$ times have
the statistical weight $w^{-n}$.
Hence, we obtain
\begin{equation}
 F = \sum_{\rm particles}w^{-n} ~~,
\end{equation}
where summation is taken for all (split) particles which have a
momentum between $u$ and $u+\Delta u$.
Error bars in spectra of figures~1--3 are calculated assuming Poisson statistics.
Taking into account the propagation of errors,
the statistical error $\Delta F$ for given momentum range $[u,u+\Delta u]$ is 
calculated as
\begin{equation}
(\Delta F)^2 = \sum_{\rm particles}w^{-2n} ~~.
\end{equation}
We adopt different values of 
$u_{\rm s1}$, $n_{\rm max}$ and $w$ for different runs
so as to obtain good statistics 
near the maximum momentum (see Appendix). 
Therefore, for different runs, the length of error bars
is different from each other.
For each case, the dependence of the cutoff shape of the electron spectrum
on $\beta$ is discussed below.

%%%%%%%%%%%%%%%%%%%%%%%%%%%%%%%%%%%%%%%%%%%%%%%%%%%%%%%%%%%%%%%%%%%%%%%%%%%

\subsubsection{Age-limited case:
$p_{\rm m,age}<{\rm min}\{p_{\rm m,cool}, p_{\rm m,esc}\}$
}
\label{sec:age}

For each value of $\beta$ (0.7, 1.0 and 1.5), 
we had five runs with different $t_{\rm age}$.
One can see from figure~\ref{fig:spec_age} that at lower electron momentum where
$p\ll p_{\rm m,age}$, the spectrum is well described by the 
analytical solution in the steady state,
$F(p)=p^3f(p)\propto p^{-1}$.
Furthermore, estimated values of $p_{\rm m,age}$ (see Table~1)
%%(see table~\ref{table1})
using equation~(\ref{eq:Emaxage}) agree with the simulation result.
Hence our present numerical scheme works well.

\begin{figure*}
\begin{center}
\includegraphics[width=55mm]{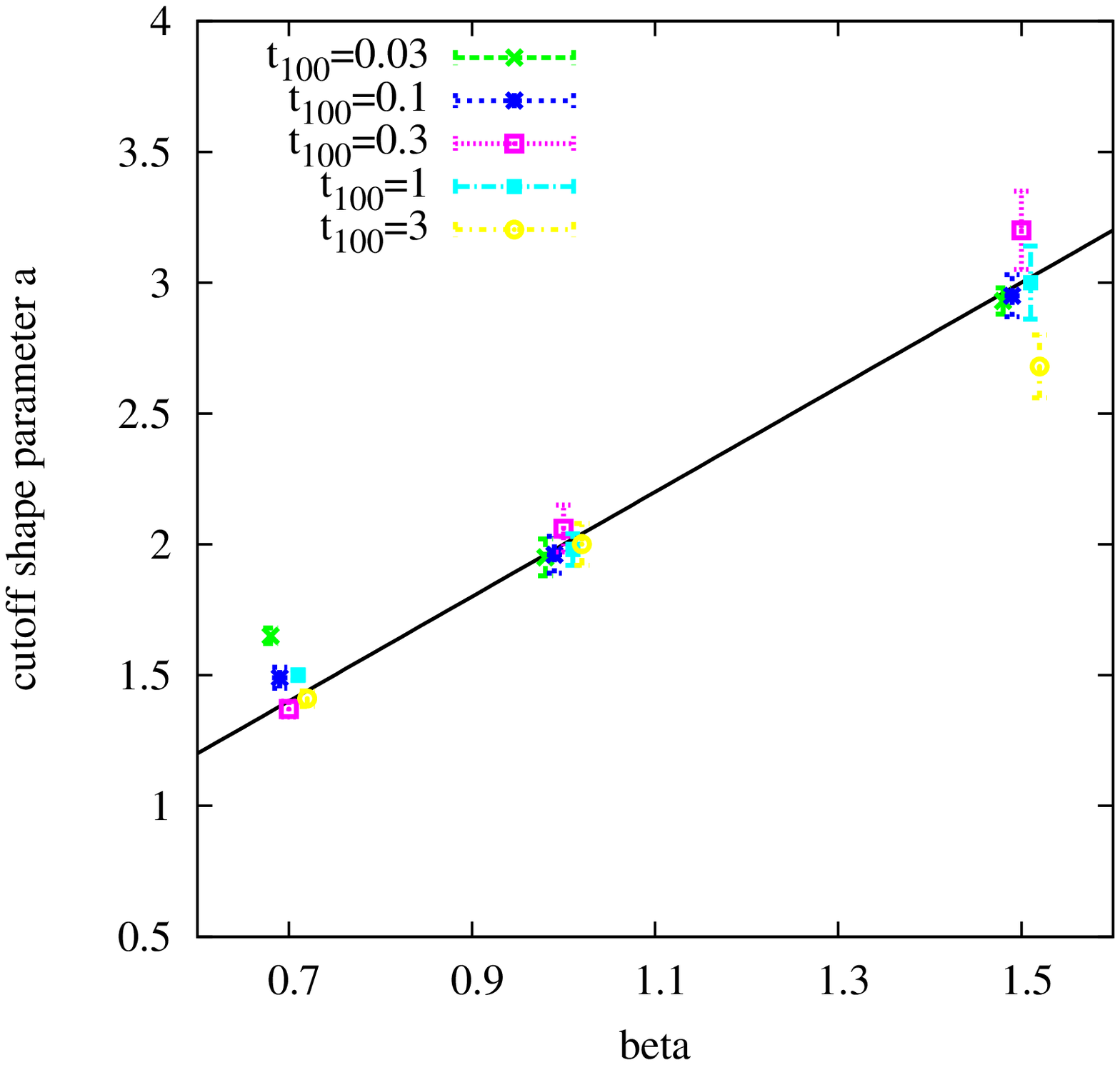}
\includegraphics[width=55mm]{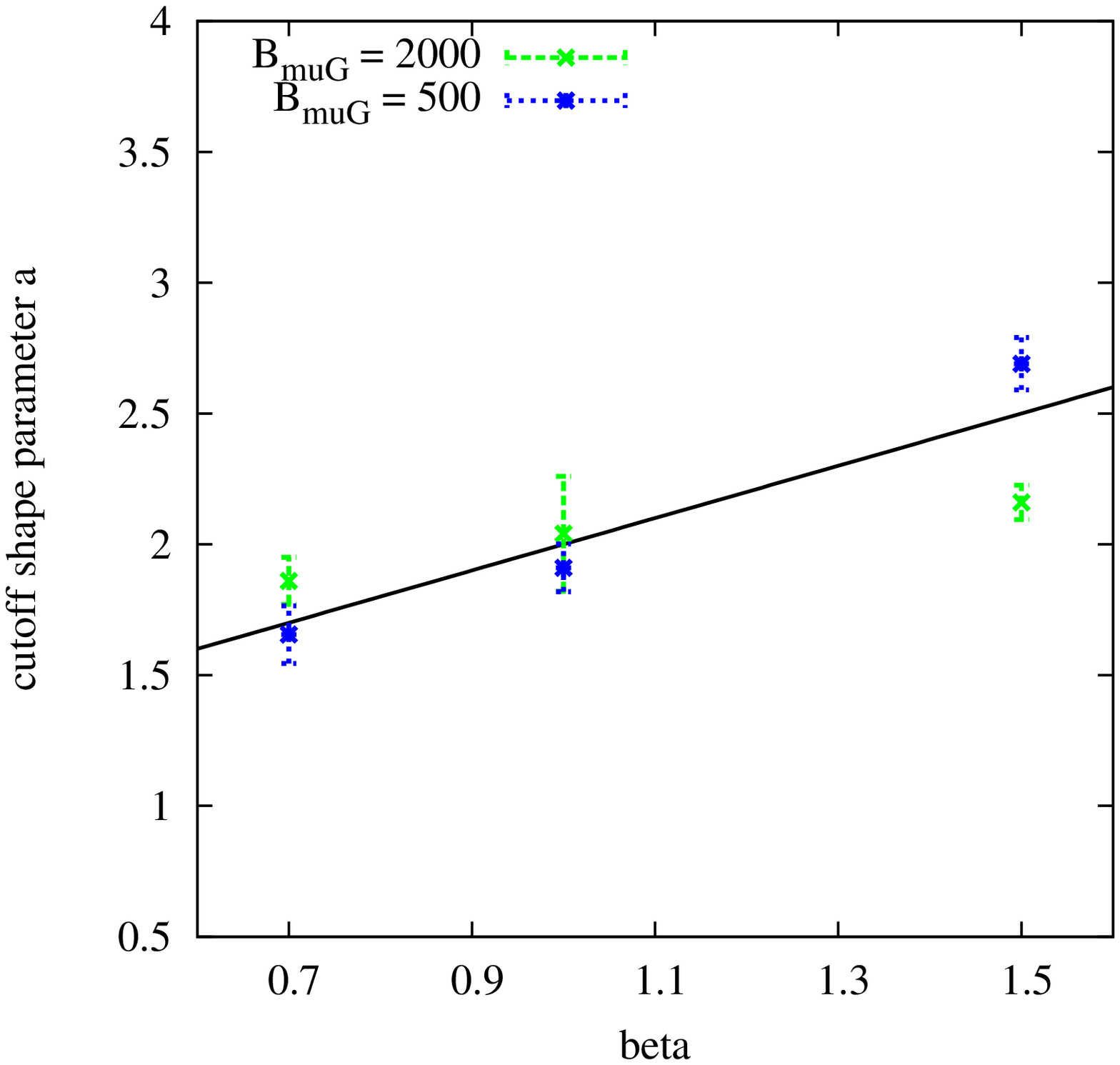}
\includegraphics[width=55mm]{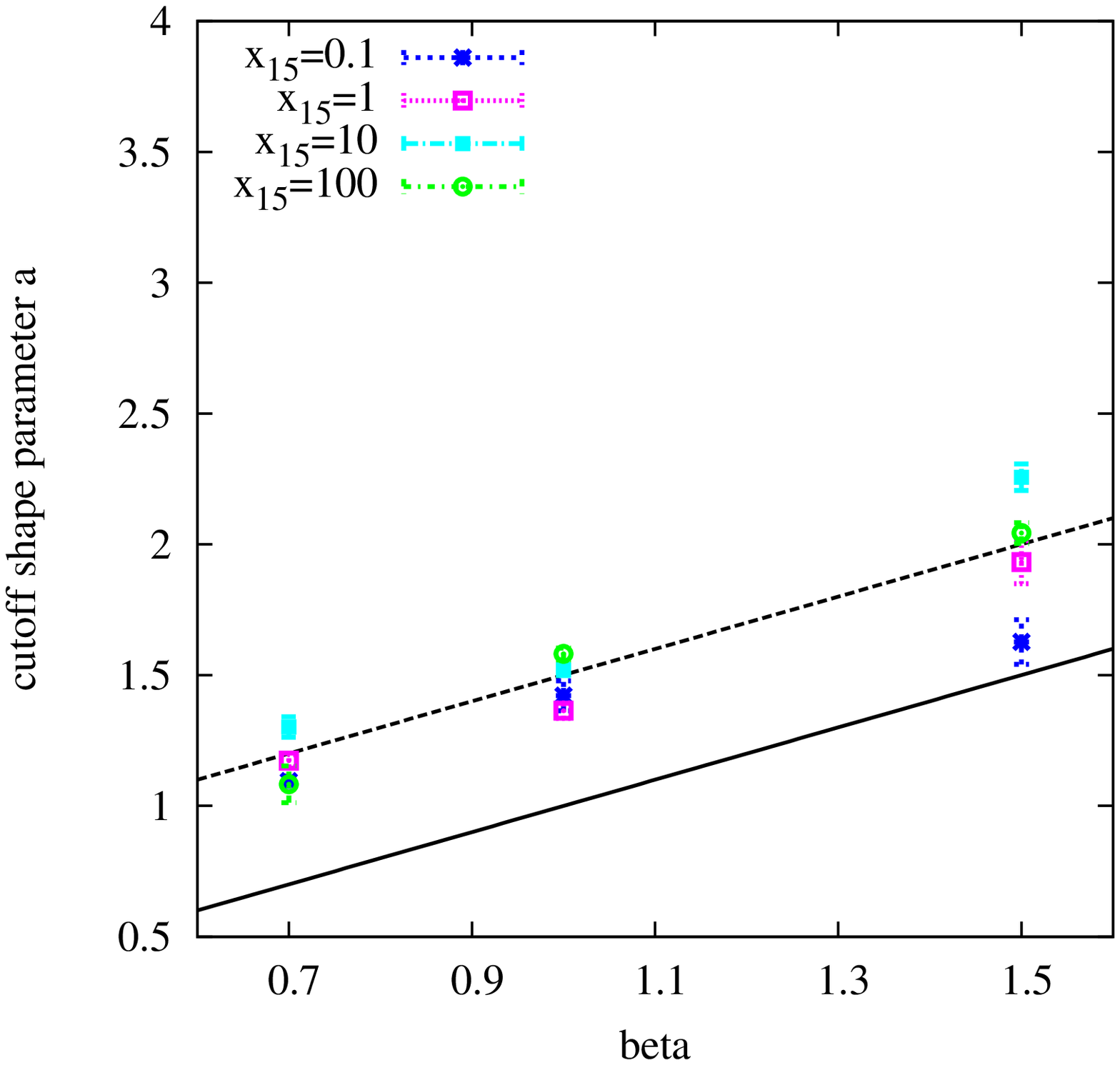}
\end{center}
\caption{
Cutoff shape parameter $a$ as a function of $\beta$ in 
the age-limited (left), cooling-limited (center) and escape-limited (right) cases.
The values of $a$ are derived by fitting the simulated spectra with
the model described by equations~(\ref{eq:shape_age}) and (\ref{eq:fit_cool})
for age-limited and cooling-limited cases, respectively.
Note that for escape-limited cases, we use phenomenological formula,
equation~(\ref{eq:shape_age}), to derive values of $a$, while we used
in Figure~3 the analytical stationary solution, equation~(\ref{eq:anal_esc}),
to fit the simulated spectra.
In the left panel, data points are artificially shifted a little in the
horizontal direction in order to be separated with each other and
to be seen clearly.
Solid lines represent $a=2\beta$, $a=\beta+1$ and $a=\beta$ for
the age-limited, cooling-limited and escape-limited cases, respectively.
The dashed line in the right panel shows $a=\beta+0.5$.
}
\label{fig:shape}
\end{figure*}

For all runs of age-limited acceleration,
we have fitted the spectrum by the following function
\begin{equation}
F(p) \propto p^{-1} \exp\left[ -\left(
\frac{p}{p_{\rm m}}\right)^a\right]~~,
\label{eq:shape_age}
\end{equation}
and obtain the value of cutoff shape parameter $a$.
The left panel of figure~\ref{fig:shape} shows the result.
One can roughly confirm earlier result of numerical simulations,
$a\approx 2\beta$ \citep{kato03,kang09}.

%%%%%%%%%%%%%%%%%%%%%%%%%%%%%%%%%%%%%%%%%%%%%%%%%%%%%%%%%%%%%%%%%%%%%%%%%%%

\subsubsection{Cooling-limited case:
$p_{\rm m,cool}<{\rm min}\{p_{\rm m,age}, p_{\rm m,esc}\}$
}
\label{sec:cooling}

For each value of $\beta$ (0.7, 1.0 and 1.5), 
we performed two simulations with the magnetic field strength
of $B=2$~mG and 0.5~mG.
Since the magnetic field strength $B$ is much
larger than any other cases, the requirement for time step,
equation~(\ref{eq:timestep}), is the most severe, so that 
we set larger $p_{\rm inj}$ of $10^4m_ec$ in order to save
the computation time.

In all runs, one can identify the cooling break,
at which the spectral slope changes from $p^{-1}$ to $p^{-2}$.
The break energy $p_{\rm b}$ is determined by the condition
$t_{\rm cool}\approx t_{\rm age}$ \citep{longair94}, 
and we derive
\begin{equation}
p_{\rm b} \approx 2.45 \times 10^{11}
B_{\mu {\rm G}}^{-2}t_{100}^{-1} m_ec~~.
\label{eq:break}
\end{equation}
For example, we obtain from this equation 
$p_{\rm b}=6.1\times10^5 m_ec$ for run C07-1,
which is consistent with the simulation result.
In some runs such as C15-1 and C15-2, 
we can see pile-ups at the high-energy end
(see the right panel of figure~\ref{fig:spec_cool}).
Based on these facts, we fit spectra which were derived from
simulations by the following function
\begin{equation}
F(p) \propto p^{-1} \, C_{\rm b}(p) \, C_{\rm p}(p) \,
\exp\left[ -\left(
\frac{p}{p_{\rm m}}\right)^a\right]~~,
\label{eq:fit_cool}
\end{equation}
where
\begin{equation}
C_{\rm b}(p) = \left[1+\left(\frac{p}{p_{\rm b}}\right)^w\right]^{-1/w}~~,
\label{eq:fit_fb}
\end{equation}
and
\begin{equation}
C_{\rm p}(p) = \left[1+\left(\frac{p}{\eta p_{\rm m}}\right)^q\right]^{k/q}~~,
\label{eq:fit_fp}
\end{equation}
describe the cooling break and the pile-up effect, respectively.
Middle panel of figure~\ref{fig:shape} shows the fitted $a$ as
a function of $\beta$.
The result is roughly consistent with the analytical result,
$a=\beta+1$, which
is derived on the steady state assumption
\citep{zirakashvili07,yamazaki13}.
In the case of $\beta=1.5$, fitted value of $a$ deviates from the analytical
expectation of 2.5. This comes from the appearance of pile-up, which deforms
the spectrum around the maximum momentum.
Hence, it is implied that the analytical expectation $a=\beta+1$ is not
always hold when the electron pile-up becomes significant.

Previously, based on similar numerical simulation to ours, 
\citet{marcowith10} numerically obtain the shock front energy spectra,
$F_0(p)=F(x=0,p)$,
for the cases of $\beta=0$, 1/2 and 1, and confirmed the relation
$a=\beta+1$ (for smaller dynamic range of momentum than ours).
In the present study, we find that at least $\beta<1$,
the relation $a=\beta+1$ roughly holds even for the 
spectrum of the whole region.

%%%%%%%%%%%%%%%%%%%%%%%%%%%%%%%%%%%%%%%%%%%%%%%%%%%%%%%%%%%%%%%%%%%%%%%%%%%

\subsubsection{Escape-limited case:
$p_{\rm m,esc}<{\rm min}\{p_{\rm m,age}, p_{\rm m,cool}\}$
}
\label{sec:escape}

In this case, steady-state spectrum at the shock front ($x=0$)
has been analytically derived
as \citep{caprioli09b,reville09,yamazaki13}
\begin{eqnarray}
F_0(p)&=&F(x=0,p) \nonumber\\
&\propto& p^{3} \exp\left[
-\frac{4}{\beta}\int^{y(p)}\frac{{\rm d}\,\log y}{1-e^{-1/y}}
\right]~~,
\label{eq:anal_esc}
\end{eqnarray}
where $y(p)=(p/p_{\rm m})^\beta$.
For each value of $\beta$ (0.7, 1.0 and 1.5), 
we have four runs with different $x_{\rm feb}$.
Then, we find that in all runs, the derived spectra are well
fitted with models described by equation~(\ref{eq:anal_esc})
(see figure~\ref{fig:spec_esc}).
Hence, equation~(\ref{eq:anal_esc}) well reproduces the spectrum
of the whole region with good accuracy although it has been
derived as the shock front spectrum at $x=0$.

Here we discuss on whether the simulated spectra are fitted with 
phenomenological formula, equation~(\ref{eq:shape_age}).
In the limit $p\gg p_{\rm m}$, we can approximate 
$1-e^{-1/y}\approx 1/y$  in equation~(\ref{eq:anal_esc}), resulting in
$F_0(p)\propto \exp[-(p/p_{\rm m})^\beta]$, 
so that one might expect $a=\beta$ \citep{yamazaki13}.
If we fit numerically derived spectra
with equation~(\ref{eq:shape_age}),
then the fitted values of $a$ are not along with the expectation
$a=\beta$.
Difference between equations~(\ref{eq:anal_esc}) and
(\ref{eq:shape_age}) with $a=\beta$ is large at $p\sim p_{\rm m}$.
However, they  seem to lie on $a\approx\beta+0.5$
(see the right panel of figure~\ref{fig:shape}). 
This implies that 
$F(p)\propto p^{-1}\exp[-(p/p_{\rm m})^\beta]$ is not 
a good approximation 
around the maximum momentum $p_{\rm m}$.

%%%%%%%%%%%%%%%%%%%%%%%%%%%%%%%%%%%%%%%%%%%%%%%%%%%%%%%%%%%%%%%%%%%%%%%%%%%

\section{Summary and Discussion}
\label{sec:summary}

We have developed a numerical method of SDE to simulate electron acceleration
at astrophysical shocks.
Our code involves Zhang's method of skew Brownian motion and particle splitting.
Using this code, we have performed simulations of electron acceleration at
stationary plane parallel shock, and
we reproduced the analytical result in the momentum range
much smaller than the maximum momentum 
--- $f(p)\propto p^{-4}$ in the age-limited and escape-limited cases,
and the broken power-law which changes from $f(p)\propto p^{-4}$ to $p^{-5}$ 
at the cooling break in the cooling-limited cases.
These results can be achieved due to incorporation of Zhang's method.
Furthermore, 
the maximum electron momentum in the simulated spectra
can be well explained by simple analytical argument, which is 
the outcome of the particle splitting method.
Therefore, we believe our numerical code works well, and it 
enables us to study the cutoff shape of the electron spectrum.

We have performed simulations for various parameter sets, and studied
how the cutoff shape, which is characterized by cutoff shape parameter $a$,
changes with the momentum dependence of the diffusion coefficient $\beta$.
In the age-limited cases, we have reproduced previous results of other
authors, $a\approx2\beta$. 
In the cooling-limited cases,
the analytical expectation $a\approx\beta+1$ is roughly reproduced
although we recognize deviations to some extent (runs~C15-1 and C15-2)
when the pile-up effect is significant.
However, we have found in the Bohm type diffusion, $K\propto p$ and $\beta=1$, 
the cutoff shape parameter $a$ is consistent with the analytical prediction
$a=2.0$ {\it both} in the age-limited and cooling-limited cases.
Hence, if the effect of escape can be neglected, 
$a=2$ should be canonical value.
Note that in the present study, we have assumed 
plane shock geometry and 
constant electron injection.
In reality, the SNR shock is nearly spherical although it has
fluctuation \citep[e.g.,][]{inoue12}.
In the spherical shock case, 
accelerated particles downstream of the shock experience
adiabatic losses. This reduces the mean energy gain they
experience at the shock, which steepens the spectral slope
\citep[e.g.,][]{yamazaki06,schure10}.
In addition, the exact spectral slope
in a time-dependent calculation depends on the injection history,
which may be complicated depending on the 
shock velocity, ambient density and so on.
These effects may influence the electron spectrum.
However, we are currently interested in the energy region near the upper
end of the spectrum.
At the given epoch, the spectrum around the maximum energy is
dominated by those which are being accelerated at that time because 
previously accelerated particles have suffered the adiabatic losses 
during transported downstream of the shock.
Hence, one can expect that the cutoff shape of the spectrum does not so much
depend on the past acceleration history \citep{yamazaki06,yamazaki13}.
This issue has not yet been studied in detail except for a few works
in which the spherical and planar shock cases were compared
\citep{schure10,kang15},
and should be investigated more in future works.

The maximum momentum is sometimes determined by the escape of
accelerated particles upstream.
In this case, we should use 
the functional form given by equation~(\ref{eq:anal_esc}), 
otherwise we should use
(\ref{eq:shape_age}) with $a=\beta +0.5$.
Electron acceleration at SNRs is sometimes limited by the escape
\citep[see figure~1 and 2 of][]{ohira12},
as well as proton acceleration, which might be inferred by
recent gamma-ray observation \citep[e.g.,][]{ohira10}.
In the present study we adopt weak magnetic field in all runs of 
the escape-limited cases, so that synchrotron cooling effect can
be neglected. Hence, our result for the escape-limited cases
is applicable to the proton acceleration.
The cutoff shape around the maximum proton momentum may be
studied by the precise gamma-ray spectrum which will be taken
in the near future.

In the present study, we have used the test particle approximation,
neglecting feedback of the accelerated particles on to the plasma
forming background shock structure.
Not electrons but protons deform the background plasma, because
they are coupled with each other through the waves excited by
accelerated protons themselves.
Various authors focus on the feedback processes of accelerated
particles on to the magnetohydrodynamic properties around the shock.
\citep[e.g.,][]{berezhko99,malkov01,kang05,vladimirov06,terasawa07,
caprioli09a,yamazaki09,zirakashvili12,bykov14}.
In such cosmic-ray modified shocks, electron acceleration is
also affected by the shock deformation,
and the results may be different from those
in the test particle limit.
These studies are remained as a future work.

%%%%%%%%%%%%%%%%%%%%%%%%%%%%%%%%%%%%%%%%%%%%%%%%%%%%%%%%%%%%%%%%%%%%%%%%%%

\section*{Acknowledgments}

Some numerical computations in this work were carried out 
at the Yukawa Institute Computer Facility.
The authors wish to thank Tsunehiko~Kato, Kohta~Murase, Aya~Bamba,
Kazunori~Kohri, Toshio~Terasawa, Fumio~Takahara
and Hajime~Takami for useful comments and discussions. 
We also thank anonymous referee for valuable comments to improve the paper.
This work was supported in part by the fund from Research Institute, 
Aoyama Gakuin University (R.~Y.), 
and by grant-in-aid from the Ministry of Education, Culture, Sports, 
Science, and Technology (MEXT) of Japan, 
No.~24$\cdot$8344 (Y.~O.), No.~22540264 (S.~Y.).

%%%%%%%%%%%%%%%%%%%%%%%%%%%%%%%%%%%%%%%%%%%%%%%%%%%%%%%%%%%%%%%%%%%%%%%%%%%

\section*{Appendix: Parameters for numerical simulation}

Parameters for numerical simulation are summarized in Table~2.
Time step, $\Delta t$, must satisfy the condition~(\ref{eq:timestep}).
In the present study, it is taken to be smaller than $2K_1(p_{\rm inj})/v_1^2$.

We need four parameters, $u_{\rm s0}$, $u_{\rm s1}$, $n_{\rm max}$
and $w$, to carry out particle splitting.
In the present study, we set
$u_{\rm s0}=\ln(p_{\rm inj}/m_ec)$.
The other parameters are taken differently for each run.

\begin{table}
\label{table2}
\centering
\begin{minipage}{80mm}
\caption{Adopted parameters in the present study.}
\begin{tabular}{@{}ccccc@{}}
\hline
Run\footnote{See Table~1.} & 
$\Delta t$ &    $n_{\rm max}$        &    $w$      & 
$p_{\rm s1}$\footnote{$p_{\rm s1}/m_ec=\exp[u_{s1}]$.} \\
&  [$10^3$s] & & &  [$m_ec$] \\
\hline
\hline
A07-1  & 10   &  6 & 12 & $10^{9.0}$ \\
A07-2  & 10   &  6 & 12 & $10^{9.0}$ \\
A07-3  & 10   &  6 & 12 & $10^{9.0}$ \\ 
A07-4  & 10   &  6 & 12 & $10^{9.0}$ \\ 
A07-5  & 10   &  6 & 12 & $10^{9.0}$ \\  
\hline
A10-1  & 10   &  6 & 12 & $10^{8.0}$ \\
A10-2  & 10   &  6 & 12 & $10^{8.0}$ \\
A10-3  & 10   &  6 & 12 & $10^{8.0}$ \\ 
A10-4  & 10   &  6 & 12 & $10^{8.0}$ \\
A10-5  & 10   &  6 & 12 & $10^{8.0}$ \\
\hline
A15-1  & 1    &  6 & 12 & $10^{9.0}$ \\
A15-2  & 1    &  6 & 12 & $10^{9.0}$ \\
A15-3  & 1    &  6 & 12 & $10^{9.0}$ \\
A15-4  & 1    &  6 & 12 & $10^{9.0}$ \\
A15-5  & 1    &  6 & 12 & $10^{9.0}$ \\
\hline
\hline
C07-1  & 5    &  5 & 10 & $10^{6.5}$ \\ 
C07-2  & 8    &  5 & 10 & $10^{6.5}$ \\
\hline
C10-1  & 0.5  &  5 & 10 & $10^{6.5}$ \\
C10-2  & 1    &  5 & 10 & $10^{7.0}$ \\
\hline
C15-1  & 0.8  &  5 & 10 & $10^{6.5}$ \\
C15-2  & 3.2  &  5 & 10 & $10^{6.5}$ \\
\hline
\hline
E07-1  & 0.11 &  6 & 10 & $10^{6.0}$ \\
E07-2  & 0.11 &  6 & 10 & $10^{6.0}$ \\
E07-3  & 0.11 &  6 & 10 & $10^{7.0}$ \\
E07-4  & 0.11 &  6 & 10 & $10^{7.0}$ \\
\hline
E10-1  & 0.11 &  6 & 10 & $10^{6.0}$ \\
E10-2  & 0.11 &  6 & 10 & $10^{6.0}$ \\
E10-3  & 0.11 &  6 & 10 & $10^{6.0}$ \\
E10-4  & 0.11 &  6 & 10 & $10^{6.0}$ \\
\hline
E15-1  & 0.11 &  5 & 4  & $10^{2.5}$ \\
E15-2  & 0.11 &  6 & 10 & $10^{5.5}$ \\
E15-3  & 0.11 &  6 & 10 & $10^{5.5}$ \\
E15-4  & 0.11 &  6 & 10 & $10^{5.5}$ \\
\hline
\end{tabular}
\end{minipage}
\end{table}

%%%%%%%%%%%%%%%%%%%%%%%%%%%%%%%%%%%%%%%%%%%%%%%%%%%%%%%%%%%%%%%%%%%%%%%%%%%

%% References
%%
%% Following citation commands can be used in the body text:
%% Usage of \cite is as follows:
%%   \cite{key}          ==>>  [#]
%%   \cite[chap. 2]{key} ==>>  [#, chap. 2]
%%   \citet{key}         ==>>  Author [#]

%% References with bibTeX database:

%% \bibliographystyle{model1-num-names}
%% \bibliography{<your-bib-database>}

%% Authors are advised to submit their bibtex database files. They are
%% requested to list a bibtex style file in the manuscript if they do
%% not want to use model1-num-names.bst.

%% References without bibTeX database:

\end{document}